# CLUSTERING OF DIFFUSE INFRARED BACKGROUND LIGHT

A. Kashlinsky
*NORDITA, Blegdamsvej 17, Copenhagen, DK-2100 Denmark*
J.C. Mather
*Code 685, Goddard Space Flight Center, Greenbelt, MD 20771 USA*
S. Odenwald
*Applied Research Corporation and Code 685, Goddard Space Flight Center*
M. Hauser
*Space Telescope Science Institute, Baltimore, MD 21218 USA*

**Abstract**

We outline a new method for estimating the cosmic infrared background using the spatial and spectral correlation properties of infrared maps. The cosmic infrared background from galaxies should have a minimum fluctuation of the order of 10% on angular scales probed by the DIRBE (Diffuse Infrared Background Experiment) on board COBE (Cosmic Background Explorer). We present the theoretical basis for the analysis and discuss the all-sky all-wavelengths survey that we conducted. We further show that a linear combination of maps at different wavelengths can greatly reduce the fluctuations produced by foreground stars in the near-IR, while not eliminating the fluctuations of the background from high redshift galaxies. The method is potentially very powerful, especially at wavelengths in the mid- to far-IR where the foreground is bright but smooth.

## 1. Introduction

Diffuse background radiation fields contain information about the entire history of the universe, including those periods in which no discrete objects exist or can be detected by telescopic study. The cosmic infrared background (CIB) should contain much of the luminosity of the early generations of stars and galaxies, since the cosmic expansion redshifts the wavelength of peak luminosity from these sources from the UV and visible bands into the infrared. In addition, dust absorbs much of the original UV luminosity and re-emits it in the IR. The CIB therefore contains important information about the evolution of the Universe between the redshift of last scattering (probed by the microwave background) and today (probed by optical surveys).

Numerous models have been developed to predict the expected properties of the CIB

at wavelengths from the near-IR ($\lambda \sim$1-10 $\mu$m) to the far-IR (e.g. Bond, Carr and Hogan 1986; Beichman and Helou 1991; Franceschini *et al.* 1991). The spectral properties and the amplitude of the predicted CIB depend on the various cosmological assumptions like the cosmological density parameter, the history of star formation, and the power spectrum of the primordial density field among others. A typical prediction over a range of wavelengths probed by DIRBE, 1.25-240 $\mu$m, is $\lambda I_\lambda \sim 10$ nWm$^{-2}$sr$^{-1}$ (e.g. Pagel 1993 and references therein). The precise measurement of the CIB levels at various wavelengths would provide information about the various cosmological parameters and scenarios. At most wavelengths 10 nWm$^{-2}$sr$^{-1}$ is considerably lower than the residual foreground levels left after the direct foreground modelling methods (Hauser 1995a,b), especially in the mid-IR where zodiacal emission is very bright. The DIRBE limits from direct modelling (Hauser 1995) have already placed interesting limits on the more exotic models of the early Universe. At wavelengths $\lambda = (15 - 60)\mu$m there exists an interesting possibility of an indirect detection from $\gamma$-rays at the level of $\lambda I_\lambda \simeq 6h(\frac{\lambda}{\mu\text{m}})^{0.55}$ nWm$^{-2}$sr$^{-1}$ (Dwek and Slavin 1994). Because of the difficulty of direct foreground modelling and subtraction, it is also important to try alternative methods as well.

One such method can be based on the angular correlation function analysis of the DIRBE maps, the groundwork for which has been laid in Kashlinsky, Mather, Odenwald and Hauser (1995, 1996; hereafter Paper I, Paper II). Here we review the theoretical background for the method and report on the limits it sets on the CIB.

## 2. Theory

The correlation signal for the CIB has been studied for a variety of cosmological models (e.g. Bond, Carr and Hogan 1986, Wang 1991, Coles, Treyer and Silk 1990, Paper I). In general the intrinsic correlation function of the diffuse background, $C(\theta) \equiv \langle \lambda \delta I_\lambda(\mathbf{x}) \cdot \lambda \delta I_\lambda(\mathbf{x} + \theta) \rangle$ with $\delta I \equiv I - \langle I \rangle$, produced by a population of emitters (e.g. galaxies) clustered with a 3-dimensional correlation function $\xi(r)$ is given in the small angle limit ($\theta << 1$) by:

$$C(\theta) = \int_0^\infty A_\theta(z)(\frac{d\lambda I_\lambda}{dz})^2 [\Psi^2(z)(1+z)^2\sqrt{1+\Omega z}]dz \tag{1}$$

where it was assumed that the cosmological constant is zero, $\Psi(z)$ is the factor accounting for the evolution of the clustering pattern, and

$$A_\theta(z) = 2R_H^{-1} \int_0^\infty \xi(\sqrt{v^2 + \frac{x^2(z)\theta^2}{(1+z)^2}})dv \tag{2}$$

Here $R_H = cH_0^{-1}$, and $x(z)$ is the comoving distance. After convolving (1) with the beam the zero-lag signal for the DIRBE sky becomes

$$C_\vartheta(0) = \int_0^\infty A_\vartheta(z)(\frac{d\lambda I_\lambda}{dz})^2 [\Psi^2(z)(1+z)^2\sqrt{1+\Omega z}]dz, \tag{3}$$

$$A_\vartheta(z) = \frac{1}{2\pi R_H} \int_0^\infty P_{3,0}(k)kW(\frac{kx(z)\vartheta}{1+z})dk. \tag{4}$$

For the top-hat beam of DIRBE the window function is $W(x) = [2J_1(x)/x]^2$ and $\vartheta$=0.46°. In the above equation $P_{3,0}(k)$ is the spectrum of galaxy clustering at the present epoch. If $\xi$ is known the measurement of $C(\theta)$ can give information on the diffuse background due to material clustered like galaxies (e.g. Gunn 1965, Peebles 1980). Such a method was successfully applied in the V (Shectman 1973,1974) and UV bands (Martin and Bowyer 1989). In the rest of the paper we omit the subscript $\vartheta$ in (3) with $C(0)$ referring to the DIRBE convolved zero-lag correlation signal, or the variance of the map $\langle (\lambda \delta I_\lambda)^2 \rangle$.

As was shown in Paper I for scales dominating the integral in (2) the 2-point correlation function at the present epoch can be approximated as a power-law. In that case in the Friedman-Robertson-Walker Universe $A_\vartheta(z)$ would reach a minimum at $z \geq 1$ whose value is almost independent of $\Omega$ (Paper I). This argument was generalized in Paper II to a more realistic $\xi(r)$ which was computed using the correlation data from the APM (Maddox et al. 1990) survey with the APM power spectrum taken from 1) the inversion technique of Baugh and Efstathiou (1993) and 2) the empirical fit to the APM data on the projected galaxy correlation function from Kashlinsky (1992). Thus we can rewrite (3) as an inequality in order to derive an upper limit on a measure of the CIB flux from clustered material from any upper limit on $C(0)$ derived from the DIRBE data:

$$(\lambda I_\lambda)_{z,rms} \leq B\sqrt{C(0)} \qquad (5)$$

where $B \equiv 1/\sqrt{\min\{A_\vartheta(z)\}} = (11-13)$ over the entire range of parameters, and the measure of the CIB flux used in (5) is defined as $[(\lambda I_\lambda)_{z,rms}]^2 \equiv \int (\frac{d\lambda I_\lambda}{dz})^2 [\Psi^2(z)(1+z)^2\sqrt{1+\Omega z}]dz$. The latter is $\simeq \int (\frac{d\lambda I_\lambda}{dz})^2 dz$ since the term in the brackets has little variation with $z$ for two extremes of clustering evolution, when it is stable in either proper or comoving coordinates (Peebles 1980). Thus the CIB produced by objects clustered like galaxies should have significant fluctuations, $\sim 10\%$ of the total flux, on the angular scale subtended by the DIRBE beam. In general, as eq. (3) shows if the bulk of the CIB comes from higher redshift this would lead to smaller relative $\delta I/I$ and vice versa (cf. Wang 1991).

## 3. Data and analysis

The COBE was launched in 1989, and included the DIRBE instrument with the aim of studying the IR emission in ten wavebands from the near-IR (J,K,L bands at $\lambda$=1.25, 2.2 and 3.5 $\mu$m) to the far-IR extending to $\lambda$=240 $\mu$m. Four of the bands were designed to match the IRAS. The instrument is a photometer with an instantaneous field of view of 0.7$\times$0.7 degrees. It was designed to have instrumental noise less than 1 nWm$^{-2}$sr$^{-1}$ in the first eight bands; at the two longest wavelengths the instrumental noise is considerably higher and this would make, as discussed below, the correlation method less efficient there. We have used the map derived from the entire 41 week DIRBE data set available from the NSSDC. A parameterized model developed by the DIRBE team (Reach et al. 1995) was used to remove the time-varying zodiacal component from each weekly map (see Paper I for more details). The maps were constructed for each of the ten DIRBE wavebands. The DIRBE maps were pixelized using the quadrilateralized spherical cube as described by Chan and O'Neill (1974) and Chan and Laubscher (1976).

After the maps were constructed the sky was divided into 384 patches of 32$\times$32 pixels or $\simeq 10°\times10°$ each. Each field was cleaned of bright sources by the program developed by the DIRBE team. The flux distribution in each patch was modelled with a smooth fourth order polynomial component. Pixels with fluxes $> N_{\mathrm{cut}}$ standard deviations above the fitted model were removed along with the surrounding 8 pixels. Three values of $N_{\mathrm{cut}}$ were used: $N_{\mathrm{cut}} = 7, 5, 3.5$. Since any large-scale gradients in the emission are clearly due to the local foregrounds a third order polynomial was removed from each patch after desourcing.

Three types of foregrounds provide dominant contributions at various bands: in the near-IR bands (1.25-4.9 $\mu$m) the main foreground contribution to the zero-lag signal comes from the Galactic stars, in the mid-IR (bands 12-60 $\mu$m) from the zodiacal light emission, and in the far-IR bands (100-240 $\mu$m) the main contribution is from the cirrus clouds in the Galaxy. The structure of the near-IR maps and the distribution of fluctuations within the various fields have been discussed at length in Paper I; the histograms for these bands are shown there in Fig. 5. They are highly asymmetric with respect to $\delta I_\lambda$ and are consistent with those from simulated stellar catalogs. The distribution at mid- to far-IR bands is more symmetric and comes from a more extended emission. Furthermore, because of the extended character of the foreground emission at mid- to far-IR the decrease of the width of the histogram (another measure of $C(0)$) with $N_{\rm cut}$ is significantly less prominent than for the near-IR where stars provide the dominant foreground. Removal of the large-scale gradients does not lead to any noticeable difference in the histograms at the near-IR bands. On the other hand, at longer wavelengths the bulk of the foreground comes from extended emission, and removing large-scale gradients, which are clearly of local origin, changes the structure significantly. For the same reason there is more noticeable decrease in the measured $C(0)$ with $N_{\rm cut}$ after removing the local gradients.

In the near-IR where Galactic stars dominate the fluctuations of the foreground there is significant decrease in the minimal value of $C(0)$ with $N_{\rm cut}$. The reason for this is that the structure of the foreground is dominated by fluctuations in point sources (stars) affecting the small scale structure of the correlation function. In the near-IR the histogram distribution of $C(0)$ on the sky becomes increasingly asymmetric with decreasing $N_{\rm cut}$ and the bins with low $C(0)$ patches at higher clipping thresholds now have essentially the same $C(0)$. On the other hand, at mid- to far-IR there is very little change in the distribution of the zero-lag signal with decreasing $N_{\rm cut}$; this is due to the fact that the foreground emission is extended with little small-scale structure. The latter in turn leads to much smaller upper limits on the value of $C(0)$ at mid- to far-IR. At bands 9,10 the instrumental noise is substantially larger because bolometers are used, and the correlations analysis of the DIRBE maps leads to significantly less interesting limits at these wavelengths (140, 240 $\mu$m).

The $C(0)$ signal is very anisotropic, showing the shapes of the Galactic and Ecliptic planes, and hence most of it comes from the foregrounds. In fact, in the near-IR bands where Galactic stars dominate the foreground there is a strong correlation between the value of $C(0)$ and the Galactic latitude. At longer wavelengths the correlation with $b$ is significantly less pronounced due to appreciable (and at times dominant) contribution from the zodiacal foreground. The map of $C(0)$ on the sky clearly shows that the plane of the Galaxy dominates the maps at the near-IR (1.25-4.9 $\mu$m), although at 4.9 $\mu$m the ecliptic plane would dominate the flux maps. At the mid-IR bands, where zodiacal emission is significant, the $C(0)$ map shows both the Galactic and the ecliptic plane. At the far-IR where most of the emission comes from the cirrus, the Galactic plane again becomes prominent.

## 4. Colour subtraction method for reducing foreground

The signal from the foreground radiation may also be disentangled from that of galaxies by considering the properties of the foreground stellar emission that are different from the CIB. The extragalactic objects have somewhat different colors than the Galactic stars, and they are redshifted as well. Furthermore, galaxies of different morphological types have different stellar populations and this too should help distinguish the colors of the foreground emission from our Galaxy (type Sc) from those of more representative galaxies (E/S0, Sa, etc.) which are expected to dominate the CIB at $J$, $K$ and $L$-bands.

We applied these ideas so far only to the three near-IR DIRBE bands. The near-IR DIRBE maps exhibit a clear color with a relatively small dispersion. We define the color between any two bands, 1 and 2, as $\alpha_{12} = I_1/I_2$. Within each field the dispersion of colors is small, between 5% and 10%. The colors confirm that most of the light of the Galaxy in these bands comes from K and M giants (Arendt *et al.* 1994). This empirical property of the constancy of color found in the DIRBE near-IR maps can be used to reduce the foreground contribution to $C(0)$ by taking the scaled difference of two maps. On the other hand, as discussed above most CIB in the near-IR is expected to come from galaxies whose spectra are significantly redshifted and thus the CIB is not expected to be washed out by this procedure. These observations are the basis for the color subtraction method we present below.

There are two *independent* contributions to the measured fluctuations in the linear combinations of the DIRBE maps: the one that comes from the *total* foreground (stars, zodiacal light, and cirrus), which we denote as $C_{*,\Delta}(0)$, and the one that comes from the extragalactic background which we label $C_{g,\Delta}(0)$. We use the $\Delta$ notation to indicate that these variances refer to the differences of maps at two different wavelengths. The total $C_\Delta(0)$ is the sum of these two components. Let us assume that the Galactic and zodiacal foreground component has the mean color between any two wavelengths 1 and 2 of $S_{12} = \langle \alpha \rangle$ and the dispersion in the color is $s_{12}^2 = \langle (\delta\alpha)^2 \rangle$, where $\delta\alpha$ is the deviation of the pixel color from the mean of the pixels in the field. Assuming that the color fluctuations are uncorrelated with the intensities, the zero-lag signal due to the foreground in Band 1, $C_{*,1}(0)$, can be written as

$$C_{*,1}(0) \simeq (S_{12}^2 + s_{12}^2)C_{*,2}(0) + s_{12}^2 \langle I_{*,2} \rangle^2. \qquad (6)$$

Note that the approximations used in this equation are not required for the validity of the analysis of the maps, but are only to indicate that major improvements are achievable with color subtraction. Let us construct the quantity which depends on the parameters in two bands (we define $\delta_i \equiv I_i - \langle I_i \rangle$), $\Delta_{*,12} \equiv \delta_{*,1} - \beta\delta_{*,2}$, where the quantity $\beta$ is for now an arbitrary number which will be quantified later. The zero-lag signal in the "color-subtracted" quantity defined above would be given by

$$C_{*,\Delta}(0) \equiv \langle \Delta_{*,12}^2 \rangle = [(S_{12} - \beta)^2 + s_{12}^2]C_{*,2}(0) + s_{12}^2 \langle I_{*,2} \rangle^2. \qquad (7)$$

It is clear from the above that if we were to choose $\beta$ close to the mean color for the foreground, $\beta \simeq S_{12}$, then the largest term in (7) could be made nearly zero and the contribution of the Galactic foreground confusion noise to the zero-lag signal $C_{*,\Delta}(0)$ in the difference map would be minimized. Furthermore, if $s \ll S$, that contribution can be decreased over the one specified in (6) by a fairly substantial amount.

Measurement shows that indeed we have $s \ll S$. Thus the variance in the color-subtracted fluctuation $\Delta$ due to the confusion noise can be reduced by a substantial factor, with a minimum when $\beta = S_{12}$. The plots of $C_\Delta(0)$ for the data from all five fields as a function of $\beta$ for $J-K$, $J-L$ and $K-L$ subtraction, for maps of various regions selected from sky both before and after zodiacal light removal show that the minimum of $C_\Delta(0)$ is indeed quite deep (see Paper I for more details). Therefore, in constructing the color-subtracted quantities we use the values of $\beta$ for each field that correspond to the minimum of $C_\Delta(0)$. Unlike the single band zero-lag signal, which would correspond to $\beta = 0$, the minimum of $C_\Delta(0)$ is distributed significantly less anisotropically. This in turn leads to interesting and low limits on the near-IR CIB as discussed in Paper I.

## 5. Conclusions

As a result of this survey we obtained interesting limits on the CIB from the $C(0)$ measurements. These results with the relevant discussion will be presented in Paper II. Assuming the "canonical" number of 10% fluctuations on the scale of the DIRBE beam the results lead to interesting upper limits on the CIB fluxes from clustered material, particularly at mid-IR bands where the foreground is bright but smooth. Because of DIRBE detector noise, they are less interesting at 140 and 240 $\mu$m. At the J,K,L bands the no-evolution (Paper I) or reasonable evolution (Veeraraghavan 1995) models of the zero lag signal in the CIB produced by normal galaxies give numbers around (1.5-3, 0.5-1.5, 0.5-1.5) nWm$^{-2}$sr$^{-1}$ respectively. Our limits in band 1 (1.25 $\mu$m) are almost an order of magnitude higher, but in bands 2, 3 (2.2, 3.5 $\mu$m) they are comparable in magnitude. If there was an extra population of galaxies at early times they would contribute an additional signal (cf. e.g. Cole, Treyer and Silk 1992). Also any other sources of emission from clustered material in the early Universe in addition to normal galaxies would increase the above theoretical estimates.

The color subtraction works well to suppress the fluctuations due to the Galactic foreground. On the other hand the predicted extragalactic background fluctuations are unlikely to be removed by the color subtraction method because the redshifts give the background a different color than the foreground. The extragalactic signal would change by a different amount, and may even increase because of the combination of redshift and morphological effects. This is discussed in more detail in our Paper I. Since the Galaxy colors are typical of general stellar populations our method effectively reduces the contribution from the nearby galaxies, but because of the redshift effects on galaxies the overall zero-lag signal remains at the levels of the single band cases. Thus the color subtracted signal probes galaxies in a window of redshifts centered on an early epoch; the distant galaxies do not have a single color, and can not be removed in the process.

**Acknowledgements**

We acknowledge support from the NASA Long Term Space Astrophysics grant, and we thank the many members of the COBE DIRBE team whose support, results and work made this project possible. The National Aeronautics and Space Administration/Goddard Space Flight Center (NASA/GSFC) is responsible for the design, development, and operation of the Cosmic Background Explorer (*COBE*). GSFC is also responsible for the development of the analysis software and for the production of the mission data sets. The *COBE* program is supported by the Astrophysics Division of NASA's Office of Space Science and Applications.